\documentclass[]{spie}  
\usepackage[]{graphicx}

\title{Antenna-Coupled TES Bolometers for CMB Polarimetry}

\author{C.L. Kuo\supit{ab}, J.J. Bock\supit{ab}, G. Chattopadthyay\supit{b},
A. Goldin\supit{a}, S. Golwala\supit{a}, W. Holmes\supit{b},
K. Irwin\supit{c}, M. Kenyon\supit{b}, A.E. Lange\supit{a}, 
H.G. LeDuc\supit{b}, P. Rossinot\supit{a}, 
A. Vayonakis\supit{a}, G. Wang\supit{a}, M. Yun\supit{d} and 
J. Zmuidzinas\supit{ab} 
\skiplinehalf
\supit{a}Observational Cosmology, Mail Code: 59-33, 
California Institute of Technology, 1200 E California Blvd., 
Pasadena, CA 91125, USA;\\
\supit{b}Jet Propulsion Laboratory, 4800 Oak Grove Dr., 
Pasadena, CA 91109, USA;\\
\supit{c}National Institute of Standards And Technology, 325 Broadway, 
Boulder, CO., USA; \\
\supit{d}University of Pittsburgh, 348 Benedum Engineering Hall, 
Pittsburgh PA 15261, USA
}

\authorinfo{For further information contact Chao-Lin Kuo, e-mail: 
clkuo@astro.caltech.edu}

\begin{document} 
  \maketitle 

\begin{abstract}

We have developed a completely lithographic antenna-coupled bolometer 
for CMB polarimetry. 
The necessary components of a millimeter wave radiometer --- a beam forming 
element, a band defining filter, and the TES detectors 
--- are fabricated on a silicon chip with photolithography.
The densely populated antennas allow a very efficient use of the focal plane
area.
We have fabricated and characterized a series of prototype devices. 
We find that their properties, including the frequency and angular 
responses, are in good agreement with the theoretical expectations. 
The devices are undergoing optimization for upcoming CMB experiments. 
\end{abstract}



\section{Introduction}
\label{sect:intro}  

The primary science goals of the next generation Cosmic Microwave Background 
(CMB) polarization experiments are to produce a high fidelity E-mode power 
spectrum and to search for the B-mode polarization. A detailed E-mode spectrum 
will help constrain the cosmological parameters, and 
test crucial hypothesis in standard cosmology. The measurements of B-mode 
polarization at high multipole value ($\ell$) can be used to study the matter 
power spectrum through 
gravitational lensing. At low $\ell$, B-mode polarization can only be produced 
by gravitational waves. The detection of this signature would provide 
invaluable information on the fundamental physics in the very early Universe,
such as the energy scale of Inflation. 
To achieve these goals in the presence of the astronomical foregrounds, the 
instruments will require wide frequency coverage, a large number of sensitive 
microwave detectors, and exquisite control of systematics\cite{weiss05}. 

Bolometers can provide photon noise-limited sensitivity over a wide frequency 
range. However, the existing feedhorn-coupled micromesh bolometers
\cite{bock95,turner01,jones03} face difficulties in extending the frequency 
coverage and scaling to kilo-pixel instruments.  
The micromesh bolometers with low thermal conductance 
(both the ``spiderweb''\cite{bock95} and the ``PSB''\cite{jones03} 
type) 
have currently been demonstrated with architectures suitable for frequencies 
$\nu>$60 GHz. 
The metallic corrugated feedhorns 
used for beam collimation are heavy and expensive to implement 
in systems with thousands of pixels at cryogenic temperatures. 
Semiconductor bolometers with high impedances 
($> M\Omega$ ) pose a challenge for kilo-pixel readout. Each of the 
bolometers requires one impedance-matching JFET, which dissipates significant 
amount of electrical power. The noise contribution from standard JFET is 
prohibitively large for use in multiplexers. So far, the number of elements in 
a semiconductor bolometric receiver is limited to a few hundreds
\cite{turner01}.

\begin{figure}[tbp]
\begin{center}  
\includegraphics[scale=0.6]{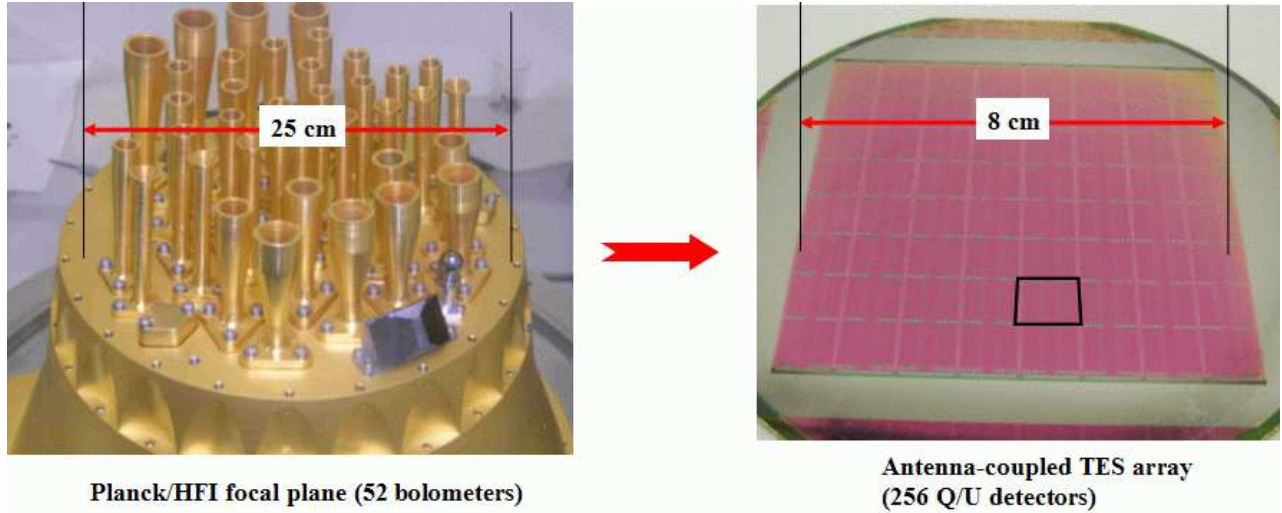}
\caption{{\em Left}: The focal plane of the ESA CMB satellite mission 
{\em Planck}, to be launched in 2008. {\em Right}: A 150 GHz prototype 
antenna-coupled bolometer array with $8\times 8$ spatial pixels and $256$ 
bolometers.
The new architecture is advantageous in weight, cost, and detector density.}
\label{focalplane}
\end{center}
\end{figure}

We describe a CMB polarimeter based on antenna-coupled superconducting 
transition edge sensors (TES) that can avoid all these difficulties.
 In this design, the beam forming slot-antenna array, microstrip 
band defining filter, and TES detectors are fabricated on a silicon chip.
These completely photolithographic detectors provide intrinsic polarization 
sensitivity and purity, collimated beams, simultaneous 
measurement of Stokes I and Q parameters in a single spatial
pixel, high focal-plane packing density, and massively parallel
fabrication. The migration from semiconductor 
bolometers to superconducting TES bolometers enables the readout of thousands 
of pixels with moderate electronics complexity. This is achieved by
superconducting quantum interference device (SQUID) multiplexers 
\cite{dekorte03,lanting05}. 
In a microstrip-coupled bolometer, only the mechanically robust slot antenna
scale with wavelength, therefore the entire frequency range ($\sim$ 30 GHz 
to 500 GHz) of interest in CMB science can be covered by the same technology. 

The densely populated antennas allow a very efficient 
use of the focal plane area. The excellent uniformity of the
photolithography provides intrinsically good alignment
and orthogonality of the two polarization modes, uniformity of the
in-line filter band-passes, and gain matching of all elements for two
orthogonal polarizations. 
Besides TES, the microwave antenna-coupled architecture is completely 
compatible with other detector technology, such as indium bump-bonded NTD 
chips and microwave kinetic inductance detectors. 

The paper is organized as follows. We describe the components of a polarization
sensitive antenna-coupled bolometer in \S\ref{sec:components}. 
The design, fabrication, and measured optical properties of 
prototype detectors are presented in \S\ref{sec:prototype}. 
Issues related to the implementation of arrays of antenna-coupled bolometers
in a polarization receiver are discussed in \S\ref{sec:system}, including a 
comparison of on-chip polarization modulators and half-wave plates 
(\S\ref{subsec:mod}). We describe our detector development plan and
several proposed CMB polarization experiments that will be using this 
technology in \S\ref{sec:plans}.

\section{ The Components of a Polarization Detector}\label{sec:components}
\begin{figure}[tbp]
\begin{center}  
\includegraphics[scale=0.8]{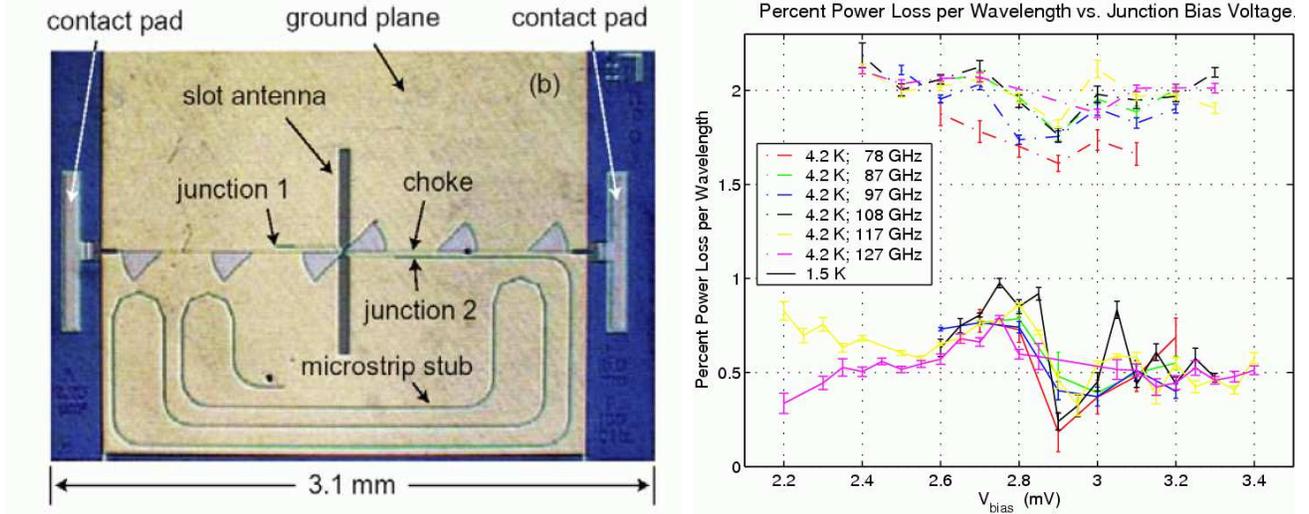}
\caption{{\em Left}: 100 GHz test device for measuring loss in
Nb/SiO/Nb. The 11.4 mm long open-ended microstrip stub sets up a periodic 
sequence of reflected nulls, detected as a function of frequency by 
sweeping the input source and measuring the power in junction 2. The
depth of the nulls depends on the transmission line loss. {\em Right}:
 Measured loss per propagation wavelength at 1.5 and 4 K. Loss per
propagation wavelength at 220 - 320 GHz and 450 - 550
GHz at 1.5 K was also measured to be ~0.5 \%.
}\label{fig:microstrip}
\end{center}
\end{figure}

\subsection{In-phase planar slot antenna} 
A beam forming element limits the radiation background on the detectors, 
and reduces side-lobes, stray light coupling, and sensitivity to cryogenic 
temperature fluctuations. Feed-coupled bolometers are especially 
advantageous compared to bare arrays under low background loading conditions 
at millimeter wave frequencies \cite{griffin02,dicker04}. In an 
antenna-coupled detector, a planar array of slot antennas perform the function
of beam collimation. The signals coming from the sub-antennas are combined 
coherently by a summing network to form a beam.

Long slots in a ground plane are intrinsically polarization sensitive since
microwave radiation tends to excite electric fields across the slots.
Another key motivation for choosing a slot architecture is that most of the 
substrate remains metalized, shielding the summing tree network from the 
incoming radiation.
We have developed 2 types of polarization selective planar antenna.
The first one is a broad-band single polarization slot antenna 
\cite{goldin03,goldin04}. In this design, parallel long slots are 
patterned in the superconducting niobium ground plane. In each slot,
a vertical microstrip summing tree with stub capacitors collects the 
incoming radiation.
A horizontal summing tree subsequently joins the vertical trees and sends 
the signal to the filter and the bolometer.  
The second design is a dual polarization antenna, 
with two colocating orthogonal sets of slots, 
and two independent microstrip summing trees and TES detectors, 
each of which measures one linear polarization.
Such dual-polarization antenna makes efficient use of the focal plane real 
estate, and reduces polarization artifacts associated with pointing errors
in certain observation modes \cite{weiss05}.

There are several research groups involved in the development of 
microstrip-coupled bolometers. Alternative beam 
collimation methods include external hyperhemispherical lenses or 
platelet feeds \cite{myers05,chuss06,weiss05}. The advantages of the 
photolithographic planar phase-antenna approach are fabrication simplicity, 
mechanical robustness at cryogenic temperatures, and immunity to misalignment.
Compared with the highly curved lenses, the flat silicon entry surface also 
greatly simplifies the anti-reflection coating. 


\subsection{The summing network}

Our array design depends critically on the loss properties of
microstrip transmission lines. We conducted measurements of the loss using the 
open-ended microstrip stub circuit, at 
75-120 GHz, 220-320 GHz and 450-550 GHz, at both 4.2 K and 1.5 K. 
A 75-120 GHz test device is shown in Fig.\ref{fig:microstrip}.
At T = 1.5 K, the loss appears to be dominated by temperature independent
loss in the dielectric (SiO). With a loss tangent ($\tan d$) 
$\sim 1.3\times 10^{-3}$, the expected loss at 150 GHz from the entire summing tree 
is on the order of 10\%. Since the loss per propagation wavelength is
nearly independent of frequency, and the pixel size scales with wavelength 
for a fixed F/number, the total loss should not change when we scale this 
design to other operating frequencies. This low loss figure is qualitatively 
confirmed in the optical efficiency measurements described in 
\S\ref{subsec:results}.

The sidelobe response of the antenna is largely determined by how each 
sub-antenna is excited by the summing network.
In all the current antenna designs, the sub-antenna arranged in a square
grid pattern are excited equally by the feed network. Consequently, the
radiation pattern exhibits minor sidelobes (at -15dB) and a four-fold symmetry.
It is straightforward to improve the symmetry and to reduce the sidelobes by 
redesigning the feed network to taper the excitation pattern. 
The microstrip feed network which collects the signals from sub-antennas
has to match the impedance of the antennas on one end, while halving 
its impedance at each binary divider. Tapered impedance transformers are 
used to maintain the downstream microstrips at a manageable width. 
We use a combination of SuperMix library\cite{ward99} and SONNET 
simulation software to optimize the design of the summing network.

\subsection{The microstrip in-line filters}
In a feedhorn-coupled bolometric radiometer, long wavelength radiation is 
rejected sharply by the waveguide cutoff, which is absent in an 
antenna-coupled bolometer. Fortunately, microstrip in-line filters can 
be easily integrated with antenna-coupled detectors. From a system 
point of view, it is in fact desirable to avoid multiple external quasioptical 
mesh filters, because they tend to generate reflections, resulting in ghost 
images. The off-axis behavior of metal mesh filters are also not 
well-studied.

We have developed a lumped-element 3rd order bandpass Chebyshev $LC$ filter,
consisting of CPW inductors and stub capacitors. This compact filter design 
does not include any ``vias'' (direct electrical contacts to the ground 
plane), and is fully compatible with photolithographic processes.
Because of their non-resonant nature, these filters do not have fundamental 
harmonic leaks. The bandgap frequency ($\sim $690 GHz) of niobium microstrips 
provides a natural high frequency cutoff for CMB experiments. In future 
applications, blocking filters are likely to be used for decreasing the 
thermal loading, while microstrip filters are used to define the science 
bands. 

\begin{figure}[tbp]
\begin{center} 
\includegraphics[scale=0.7]{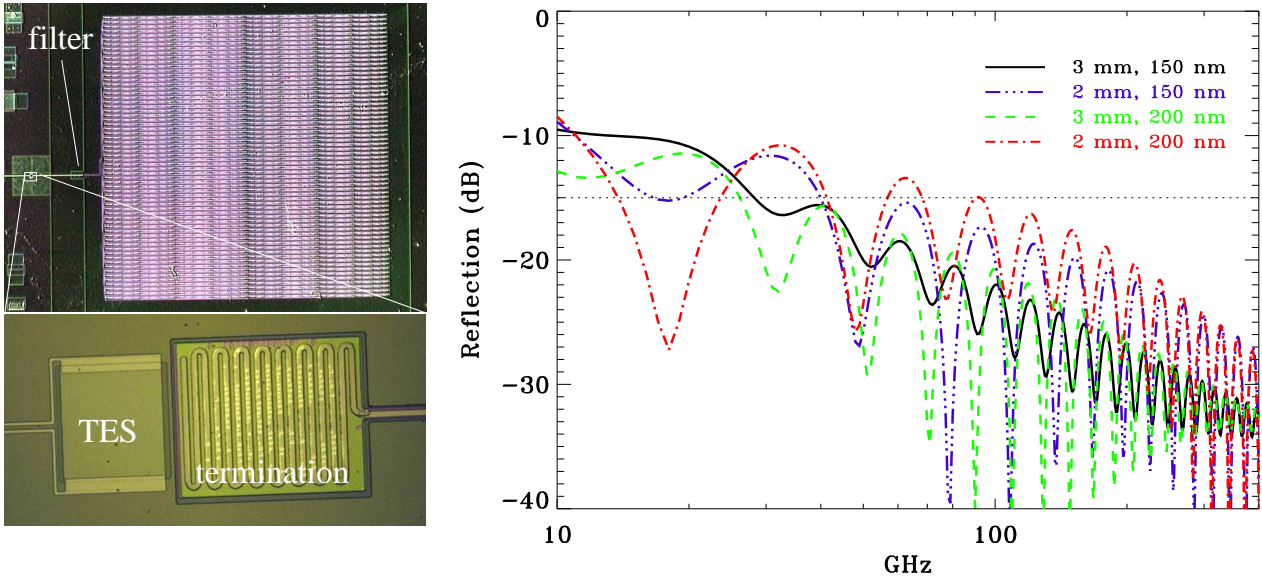}
\caption{{\em Left}: The picture of a single polarization antenna-coupled
TES bolometer. The antenna, filter, termination microstrip, and the TES film 
can be seen. {\em Right}: The residual reflection from the termination,
an open-ended 6 $\mu$m wide meandering gold microstrip transmission line. 
The calculation is performed with the SuperMix library\cite{ward99}. The 
horizontal dashed line 
indicates a 97\% efficiency (-15dB reflection). The numbers shown are 
the microstrip length and film thickness. The entire frequency range of 
CMBPol (30 GHz to 500 GHz) can be covered with this type of termination.
}\label{fig:term}
\end{center}
\end{figure}

\subsection{The TES bolometer} 
After the bandpass filter, the signal collected by the antenna is transmitted
through the Nb microstrip, and readout with microstrip-coupled
bolometers, thermally isolated on micro-machined silicon nitride beams. 
The TES sensor is coupled to a Nb/SiO/Nb microstrip, which enters the thermally 
isolated patch on a suspended silicon nitride beam, and terminates in a 
meandering normal-metal microstrip (see Fig.\ref{fig:term}). 
A TES film deposited on the isolated region, and readout via superconducting Nb
leads, detects the heat from dissipation of electromagnetic energy in the 
resistor. The bolometer operates in the standard voltage bias configuration, 
which provides strong electrothermal feedback \cite{irwin95}. The principal 
benefits of this operating mode are linearity and rapid speed of response. 

The termination resistor is made of a meander of normal metal open-ended 
microstrip. The un-absorbed EM wave is reflected at the end of the microstrip, 
as a result the effective length of the resistive microstrip is twice the 
physical length (2.2 mm).
Since the characteristic impedance is largely determined by the geometry, 
the impedances of the superconducting and resistive microstrips are 
well-matched. The termination efficiency is calculated to be $99\%$ at 
100 GHz. The advantages of this design are wide bandwidth and immunity 
to variations in the thickness and the resistivity of the lossy film.
The same meandering microstrip can be used to terminate 30 GHz radiation
with an efficiency of $92\%$. A 3 mm long microstrip 
will absorb up to $97\%$ of the incoming power (Fig.\ref{fig:term}).
The heat capacity of the gold meander is on the order of $0.1 pJ/K$
or less, small compared to that of the TES film.
The detector giving a physical time constant on the order of a few ms for 
a thermal conductance appropriate for CMB polarimetry. 
This time constant will be reduced by the factor of the loop-gain 
in the voltage biased operating condition\cite{irwin95}.
Another significant advantage of microstrip-coupled bolometer geometry is
that the thermalization time constant is much shorter than the external 
(bolometer) time constant, providing excellent thermal efficiency 
\cite{mauskopf97,turner01}. 

While in theory small-volume bolometers might someday provide 
lower noise equivalent power (NEP) at a higher physical temperature 
\cite{ali2004,karasik2000}, a micro-machined mechanically-isolated 
bolometer allows the characteristics of the termination, the thermistor, 
and the thermal conductance to be optimized independently. State-of-the-art 
photolithography can now reliably produce silicon nitride legs with
extremely high aspect ratio\cite{kenyon2005}. The NEP required by 
a space-borne CMB mission is now routinely achieved in the laboratory
with appropriate time constant.

The TES sensors will be read out using SQUID current amplifiers 
with time-domain multiplexing~\cite{dekorte03} provided by NIST.
To demonstrate the compatibility of TES and the NIST MUX, we have fabricated a 
low thermal conductance, 8 element Mo/Au TES array ($T_c\sim$ 0.12 K), and 
integrated it with NIST's SQUID multiplexing readout. The observed noise 
degradation from the readout is small: we measured a dark NEP of $3\times 
10^{-18}$ W$\sqrt{Hz}$.

Time-domain SQUID multiplexing has been under development for a number
of years and is now a fairly mature technology. 
The current SQUID MUX can multiplex up to 32 channels~\cite{dekorte03}.
In the future, the microwave frequency domain SQUID MUX ~\cite{irwin04} 
with reflectometer readout might be able to multiplex hundreds or even 
thousands of detectors, with electronics at a much lower cost. 

\section{Prototype Detectors}\label{sec:prototype}

The antenna-coupled architecture has now been fully demonstrated.
We have fabricated and thoroughly characterized 
a series of prototype antenna-coupled TES detectors. 
 We use Mo/Au (100nm/10nm) bilayers as the TES to obtain a transition 
temperature of 0.9 K, appropriate for high-background optical testing. 
In the following sections we describe the fabrication, test, and 
measured optical properties of these prototype detectors.

\subsection{Fabrication}

The devices are fabricated in the Microdevices Laboratory of the Jet 
Propulsion Laboratory.
The process begins with 200 nm of Nb sputtered onto a 100 mm diameter Si wafer 
as the ground plane. Slot antennas are then patterned using the reactive ion 
etching (RIE) process. 
In order to separate the spectral responses of the antennas and the microstrip filters, 
half of these prototype detectors are not incorporated with filters. 
Thermal SiO (400nm) was evaporated on the top of the Nb ground plane. This 
then becomes the dielectric layer for the microstrips.
The TES film dimension is 100 
$\mu$m $\times$ 95 $\mu$m, giving a normal resistance of $\sim$ 0.6 $\Omega$.
A thin Ti layer (4nm) is used to promote adhesion between the TES and the SiO 
layer.  The TES layer is then ion milled and cleaned using high power oxygen 
plasma. The absorbing resistor is a Ti/Au/Ti (4nm/150nm/4nm) meander. 
A second sputtered SiO layer (100nm) is used to protect the 
resistor layer. The wafer is then sputtered for the Nb microstrip with a 
thickness of 400nm and etched using the fine contact alignment method. 
Finally, front-side and back-side silicon nitride membranes are etched, 
followed by deep-trench Si etch process to form silicon nitride beam-isolated 
bolometers. In Fig.\ref{fig:term}, the antenna, the termination resistor, 
and the TES film (before release) are shown.

\subsection{Optical tests and results}\label{subsec:results}

\subsubsection{Test set-up} 

The devices are cooled by a closed-cycle $^3$He refrigerator.
A low pass metal mesh filter with a cut-off frequency of 280 GHz is inserted 
in the optical train for this initial testing. Carbon-loaded PTFE 
sheet with a thicknesses of 2 millimeters is used to reduce 
in-band millimeter wave loading on the detectors, providing an attenuation 
of $\sim 80\%$. Additional Fluorogold, plain PTFE sheets, and metal mesh 
filters are used for infrared blocking. A $\lambda/4$ quartz plate is glued 
onto the silicon entry surface as the anti-reflection coating. 
A superconducting niobium shield with a 4.4-cm opening is placed at 4 K 
during the measurements to guard the TES and the SQUID read-out against 
magnetic interference.

For each detector, the angular response is measured with 
an optically modulated thermal source. The frequency response is measured with a 
large aperture Fourier transform spectrometer 
(FTS). The optical efficiency (the normalization of the spectrum) is measured separately 
by illuminating the detector with a known optical power from a 
cold blackbody source in the dewar. 
It is crucial to maintain and monitor the temperature of the detector 
substrate while heating up the blackbody illuminator, since a systematic 
drift in the the bath temperature could be mis-interpreted as an optical 
power. We use a metal mesh filter and a Fluorogold sheet to filter out infrared
radiation from the illuminator. We mount an NTD thermistor chip on the detector silicon 
substrate to monitor the temperature drifts. We observe a temperature rise 
of $\sim 0.58$ mK in the substrate when the illuminator temperature is raised 
by $10$ K. This systematic drift is corrected for in the results reported here.

\subsubsection{Radiation pattern}
\begin{figure}[h]
\begin{center}  
\includegraphics[scale=0.6]{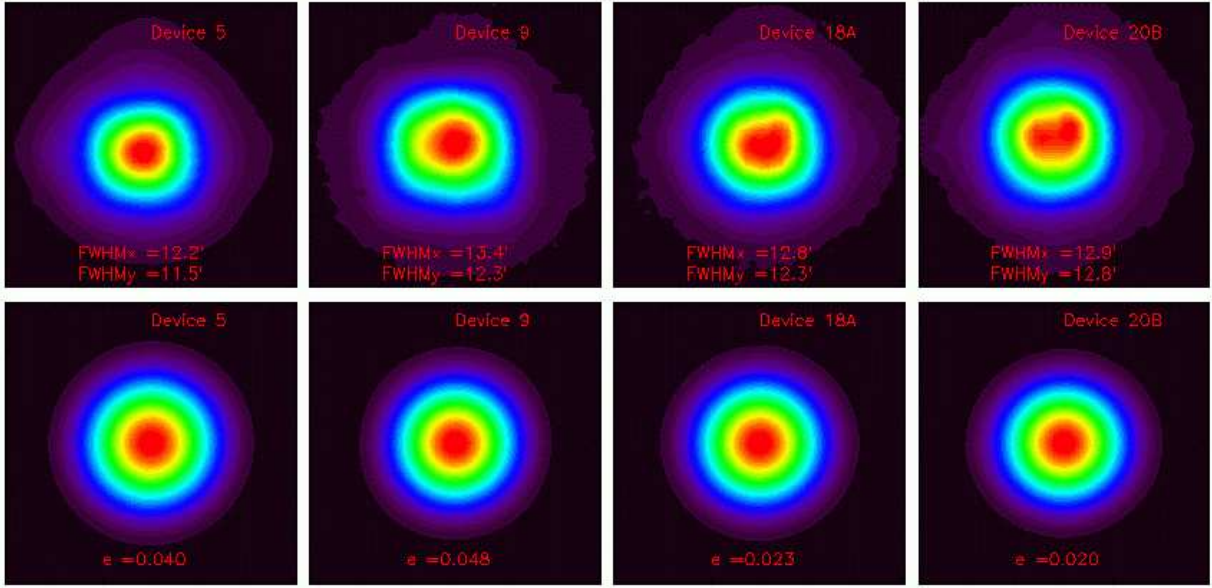}
\caption{{\em Top}: The radiation patterns of representative detectors measured 
with a chopped thermal source. From left to right are, single polarization 
antenna without the microstrip bandpass filter, single polarization with filter, 
dual polarization without filter, and dual polarization with filter.
{\em Bottom}: The Fourier transform of the radiation patterns, with a cold stop at 
a radius of 1.29 FWHM (corresponding to ``20dB'' for a Gaussian illumination). 
Neglecting distortions from geometrical optics and assuming the phase errors are small, 
they represent the beams on the sky. The number indicates the eccentricity of the beams.
These highly symmetrical beams have very low sidelobe content ($<-30$ dB).
}\label{beams}
\end{center}
\end{figure}

The measured FWHM for both the single- and dual- polarization detectors
are $\sim 13^\circ$, in good agreement with the theoretical predictions 
and previous measurements with SIS junctions using a coherent source 
\cite{goldin03,goldin04}. Out of the 8 antennas measured (4 are shown 
in Fig.\ref{beams}), 7 produce satisfactory radiation patterns. 
One dual polarization detector (not shown) exhibits a double-peak mainlobe 
and sidelobes up to 25\%. This can be explained by a broken or 
shorted microstrip in the summing tree. New fabrication equipments and techniques
(see \S\ref{subsec:fab}) will greatly improve the reliability of the summing 
network in the upcoming batch of detectors.

 We investigate the asymmetries in the radiation patterns
by repeating the measurements after a $90^\circ$ rotation of the detector. 
We find that the minor asymmetry near the peak in the radiation pattern is caused by 
the optical set-up, most likely the infrared filters at oblique angles.
The measured cross polarization responses of the antennas are also 
dominated by the reflections from these infrared filters, at
the $5\%$ level for both the single-polarization and dual-polarization 
antennas. The system is currently being re-designed to allow a more symmetric
optical path. With the present set-up, we place an upper limit on the cross-polarization
to be less than $5\%$, and the ellipticity to be less than $10\%$.

As stated earlier, the radiation pattern of a uniformly excited phased 
antenna have sidelobes at -15dB level. In the previous measurements with SIS junctions, 
these sidelobes are observed to be at the predicted level.
In our current measurements, the optical opening is not large enough for sidelobe 
characterization.
In a polarization receiver, the sidelobes will be terminated at the 
cold Lyot stop. The illumination of the pupil, and consequently the beam on the sky, 
is then determined by the main lobe of the antenna radiation pattern. Neglecting 
distortions from geometrical optics and assuming the phase errors are small, 
the beams on the sky can be approximated by 
the Fourier transform of the illumination patterns.
We take the Fourier transform of the measured radiation patterns 
(Fig.\ref{beams}) with a cold stop at a radius of 1.29 FWHM (corresponding to 
a ``20dB'' taper for a Gaussian illumination), to assess the beam shapes 
expected from a telescope fed by antenna-coupled bolometers. The resulting 
beams are highly symmetrical and have very low ($<-30$dB) sidelobe content 
(bottom rows of Figure \ref{beams}). We comment on the polarization modulation 
and beam artifacts in \S\ref{subsec:mod}.

\subsubsection{Spectral response and optical efficiency}

\begin{figure}[tbp]
\begin{center} 
\includegraphics[scale=0.8]{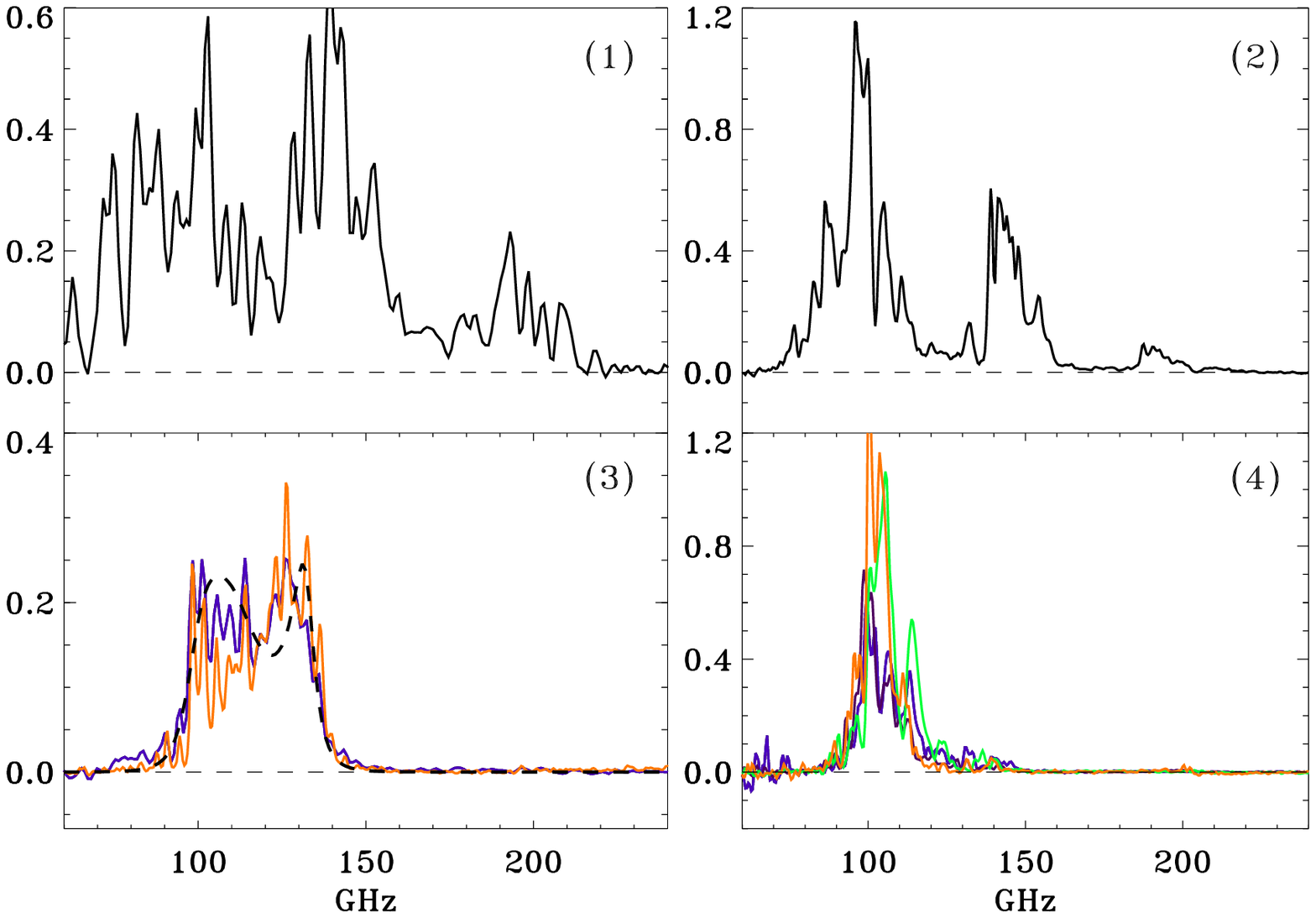}
\caption{The spectral responses of the detectors measured with a Fourier 
transform spectrometer. (1) The response of the long-slot single polarization 
antenna. (2) Dual polarization antenna.
(3) Single polarization antenna with a bandpass microstrip 
{\em LC} filter. The dashed line is a prediction of HFSS, with an adjusted dielectric constant 
(see text). (4) Dual polarization antenna with the
microstrip filter. The optical efficiency of the single polarization antennas is anomalously low
because of reflections in the summing network. This problem will be corrected in the next
detector design.
}\label{fts1}
\end{center}
\end{figure}

\begin{figure}[h]
\begin{center}  
\includegraphics[scale=0.7]{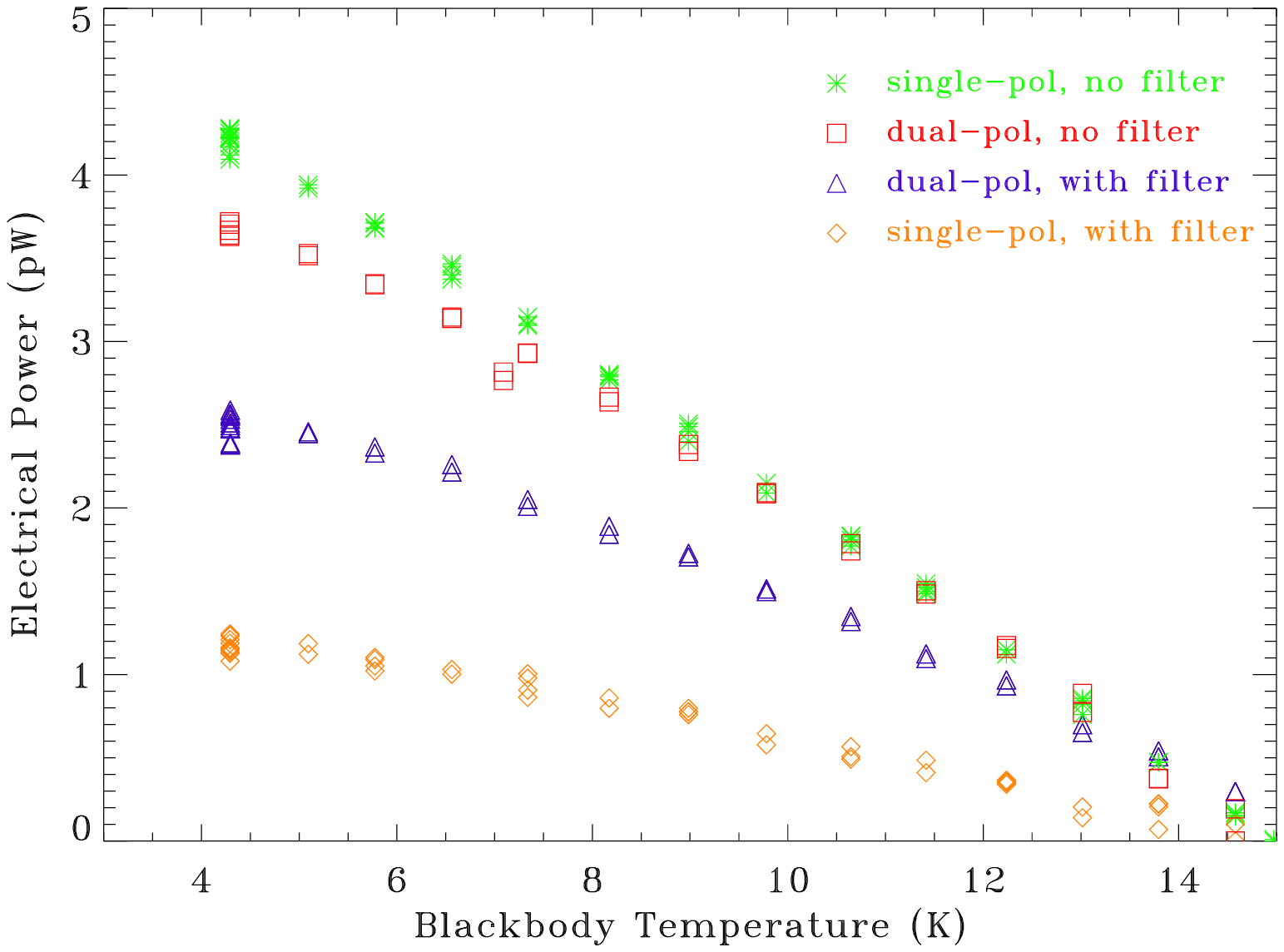}
\caption{The required electrical power to keep the TES at the transition, as a 
function of the blackbody illuminator temperature. The lines are shifted vertically for
clarity. The fitted slopes are used to normalize spectral responses shown in Figure 
(\ref{fts1}).
}\label{fig:bb}
\end{center}
\end{figure}

The FTS spectra in Figure(\ref{fts1}) are normalized with the blackbody illuminator 
measurements (Fig.\ref{fig:bb}). The effects of a Mylar beam-splitter is corrected for.
We also assume single-moded throughputs ($A\Omega$) while plotting these spectra. The 
spectrum of the unfiltered single-polarization 
device indicates that the long-slot antenna has a bandwidth in excess of $50\%$ 
(Fig.\ref{fts1}, panel 1.). The spectrum shows prominent interference fringes, with
optical efficiency lower than expected, especially near the design frequency of 100 GHz. 
We attribute the anomalously low efficiency and fringes to reflections 
in the microstrip summing network. To obtain largest possible bandwidth, we chose an 
array design with $64\times 64$ feed points. However, a more careful analysis carried out 
after the array 
fabrication indicates that the impedance transformer in the $64\times 64$ summing 
network causes much higher reflections than originally estimated. This design flaw 
can easily be fixed in the next generation devices.

The unfiltered dual-polarization antenna shows a sharp resonance at the design frequency 
of 100 GHz, as well as two additional resonances near 145 GHz and 190 GHz (Fig.\ref{fts1}, 
panel 2.). The main resonance has a peak optical efficiency around 90\%, and a 
usable bandwidth of $\approx 25\%$, agreeing with the theoretical calculations based on 
method of moments \cite{goldin03}. The $16\times 16$ summing network in the dual-polarization 
antenna incurs some reflections, but much less than that from the $64\times 64$ network. 

Comparing the spectrum of the unfiltered single-polarization detector with those of 
the filtered detectors (Figure \ref{fts1}, panel 3., spectra from two detectors are
shown), we find that the microstrip filters have a $35\%$ bandwidth, centering around $120$ 
GHz. Good uniformity is observed in the transmissions of two filters.
Overall, the microstrip filters show high 
out-of-band rejection, and sharp cut-off edges. An external high-pass thick grill filter 
is used to limit harmonic leaks between 170 GHz and 280 GHz (the cutoff of the metal mesh 
filter) to be $<0.5\%$.
The filtered dual-polarization detectors (Figure \ref{fts1}, panel 4., 
spectra from two devices, or four TES, are plotted) tell a similar story. The low frequency
end cutoff near 95 GHz is provided by the microstrip filters. The high frequency cutoff 
at $\sim$110 GHz is dominated by the antenna itself, as is apparent by comparing the two 
right panels. 

The center frequency of the microstrip filters is designed to be at 100 GHz, 20\% lower
than the measured value. To investigate the frequency shift, we measure the thickness of 
the dielectric layer with a profilometer. The difference between the 
measured thickness and the target thickness is too small to account for the 
observed disagreement. Some slight misalignment is observed in these first generation 
prototype detectors made with contact lithography. We perform numerical calculations for 
filters with misalignments in different layers (SiO, the Nb microstrip, and ground 
plane). The effects to the transmission of the filter are again found to be 
very small. We arrive at the conclusion that the dielectric 
constant is lower than the target value, presumably because of contaminations 
during the SiO deposition, resulting in a capacitance change in the {\em LC} 
circuit and the frequency shift. With an adjusted dielectric constant, we reproduce
the transmittance of the filter with numerical calculations (HFSS and SONNET). 

The blackbody illuminator in the optical efficiency measurements
is made of carbon loaded Stycast-2850 epoxy, temperature controlled between 4K and 20K.
The $T_c$ of the prototype detectors are designed to be $\sim 0.9$ K. 
We measure the thermal conductance of the bolometers, usually denoted by $G$, 
by fixing the blackbody temperature at 4K and varying the cold plate temperature.
To derive the optical efficiency, we acquire TES $I-V$ curves as a function of the
blackbody temperature while holding the cold plate temperature.  
At each blackbody temperature, several $I-V$ curves are obtained as consistency 
check. The electrical power required to maintain the TES at a fixed resistance 
(1/2 the transition) is calculated from each $I-V$ curve. 
 The thermal conductance, measured to be around 0.5 nW/K, is used to correct for 
the cold plate temperature fluctuations during TES $I-V$ curve measurements. 
After the correction, the points form a straight line, with the slope indicating
the optical coupling of the detector to the illuminator (Figure \ref{fig:bb}).
The TES normal resistance and shunt resistance are independently measured at 4K 
with a 4-wire AC bridge, minimizing the errors introduced by parasitic series 
resistance (normally $\sim m\Omega$). 

 These measurements provide normalization to the FTS spectra, however, it should 
be pointed out that the filtering in this test is 
significantly less than that in the FTS measurements. Therefore the FTS spectra 
do not exactly represent the transmittance in the optical efficiency 
measurements. 


\section{Integrating the array with a polarization receiver}\label{sec:system}

\subsection{Fabrication reliability and uniformity}\label{subsec:fab}

The summing network consists of many long and thin microstrip transmission 
lines, each of which crosses 2 steps in the ground plane created by the slot 
antenna. The prototype detectors described in \S\ref{sec:prototype}
are fabricated with contact lithography. Already, judging from the
measured angular responses, only one detector shows significant signs of 
failure in part of the summing network. The future arrays will be fabricated 
with stepper lithography, vastly improving the reliability and resolution 
of the structures. In addition, we have implemented a biased sputtered 
SiO$_2$ process. By applying a DC voltage bias of several hundred volts 
during sputtering, the SiO$_2$ layer exhibits a tapered edge 
profile, with better metal coverage and smoother, low impurity dielectric
content. To demonstrate the reliability of the step coverage, we have 
fabricated a test device in which a 5$\mu$m-wide meandering Nb microstrip 
is shown to have no breaks or shorts over its entire length of 255 mm and 
3,800 cross-overs. We believe this level of reliability is adequate for the 
production of high yield antenna-coupled TES arrays.

To reduce wiring complexity, several TES in an array will 
share a common bias\footnote{In the current design, 32 
TES using the same MUX chip will share a common bias.}. 
It is therefore important to control the transition 
temperature, the normal resistance, and the thermal conductance of each 
TES. In the strong voltage bias limit (when the shunt resistance is much 
less than the TES resistance), the electrothermal equation reads
\begin{equation}
Q\sim \frac{G_0}{(\beta+1)T_b^\beta}
(T_c^{\beta+1}-T_b^{\beta+1})-\frac{V_b^2}{R},
\end{equation}
where $Q$ is the optical power, $T_b$, $T_c$ are the bath temperature 
and the transition temperature, and $V_b$ the bias voltage. 
$G=G_0(T/T_b)^{\beta}$ defines the temperature dependent thermal conductance of the 
bolometer. A voltage-biased TES is self-biasing, in a sense that for a fixed 
bias voltage $V_b$ the resistance $R$ can vary freely from the normal 
resistance $R_n$ to zero to compensate for different loading $Q$. In practice, 
however, it is not desirable to operate the TES at a resistance below 
$R_n/3$ because of excess noise. It is apparent from this equation
that variations in $T_c$, $G_0$, and $R_n$ across an 
array degrade the dynamic range of $Q$. For millimeter/far infrared 
instruments, the dynamic range in $Q$ is usually designed to be a factor of a 
few. Moderate variations (up to 10 or 15\%) in $G_0$ and $R_n$ cause only 
minor degradation in this factor. However, since $\beta$ is usually 
in the range of $2\sim 3$ for silicon nitride at sub-K temperature, 
effects from $T_c$ variations are the most important. 

The TES in the pathfinder array for balloon-borne SPIDER\cite{montroy06} 
will be based on elemental titanium. It is known that the $T_c$ of Ti films
is close to the bulk value ($\sim$ 0.42 K) and is fairly immune to thickness
variations. To produce Mo/Cu or Mo/Au bi-layers with highly 
controlled $T_c$ poses a greater fabrication challenge. A dual-$T_c$ TES 
(0.42 K and 1.2 K) made from Ti and Al films is also under study.
Such TES can will greatly help laboratory testing under high optical
loading.

\subsection{Polarization modulation/optics}\label{subsec:mod}

Polarization modulators are often incorporated to aid the detection of a
weak signal on top of a large unpolarized background.
The types of polarization modulators include switches on the 
focal plane and external polarization modulators. The focal plane modulators
 include PIN diodes in correlation polarimeters\cite{barkats05},
Faraday rotation modulators\cite{keating03}, MEM stripline 
switches\cite{chuss06},
and superconductor-insulator-superconductor (SIS) inductance 
switches\footnote{We have successfully demonstrated an SIS switch that provides
80\% modulation over a frequency range of 10\%.}.
The most important example of an external modulator is a half-wave plate 
\cite{oxley04,montroy06}.

To compare these different approaches, we recall the two primary purposes 
of modulators. First, they can put the signal frequency above the 1/f noise
frequency. For correlation polarimeters \cite{barkats05}, a fast modulator 
is essential to avoid gain fluctuations in the amplifiers. 
Bolometers are intrinsically very stable detectors. 1/f noise lower than
0.1 Hz is routinely achieved in various bolometric CMB experiments.
Therefore, the performance in the second function of a polarization modulator, 
i.e., its ability to reduce polarization artifacts from the instrument, 
most notably the beam mismatch, passband mismatch, and gain mismatch 
\cite{weiss05, hu2003}, is what should be weighted more in evaluating 
different modulation schemes.

Measuring polarization with bolometers generally requires taking 
differences of measurements at different polarization 
angles. Many effects associated with the beam, such as 
differential ellipticity and differential pointing, can create an artificial 
polarization signal. For a B-mode detection at T/S=0.01, the requirements for 
beams are very stringent 
\cite{hu2003}. However, if the polarization differences can be 
made with the same beam, these problems can be avoided.
This idealized polarization rotation is most closely approximated by a 
rotating half-wave plate at the pupil of the optics. This is because that a 
half-wave plate locally rotates the direction of polarization without altering 
the field distribution, or the illumination. Other methods of polarization 
rotation, including sky/instrument rotation and all the on-chip modulators, 
do not help mitigate, and sometimes can even exacerbate the beam
mismatch. For a detailed account of the systematic effects associated 
with polarization measurements and how a half wave plate 
help mitigate them, see the SPIDER paper in these {\em Proceedings}
\cite{montroy06}.

On-focal plane modulators do help alleviate 1/f fluctuations, and effects 
associated with focal plane temperature fluctuations and other scan-synchronous
pick-ups. Given their advantages 
in speed, weight, and mechanical and cryogenic simplicity, on-focal plane 
modulators are still very worth pursuing alongside with the half-wave plates. 
For the first generation antenna-coupled TES, however, we assume that a
rotating half-wave plate at the pupil is the primary mode of modulation.
With the half-wave plate, the requirements for the beam are much more relaxed. 
The beams should be symmetric enough to 
produce a window function that is isotropic in $\vec{\ell}$ 
space, and low in sidelobe content to avoid astronomical foregrounds.
The beams produced by antenna-coupled bolometers (Figure \ref{beams}) have already
met these requirements.  

\section{Future plans}\label{sec:plans}

The current antenna-coupled TES prototype devices provide the first successful 
demonstration of this promising technology. We will continue to refine the microwave 
designs, with help from rapid feedback of optical characterizations. The wide 
bandwidth and well-understood impedance properties of long slot single polarization 
antenna can be used to study various downstream components, such as the summing 
network and filters. We have designed a new dual polarization architecture with a much 
larger bandwidth than that of the current design. The new 
antenna will be implemented in the next generation test detectors. Our current lumped-element
filter design offers very good results in terms of in-band transmittance and out-of-band 
rejection. The absolute frequency control should improve with better fabrication 
techniques and experience.

Magnetic field shielding is an important issue for TES array and SQUID readout.
We are currently setting up a SQUID multiplexing system 
for an adiabatic demagnetization refrigerator (ADR) system to test different
magnetic field shielding schemes. In particular, we have the option of individually 
shielding each SQUID MUX chip with a superconducting enclosure. 
It is also necessary to shield the strong fields from the ADR. From our 
experience, an attenuation of 60-80 dB can be achieved
with a few layers of high permeability cryoperm shield, Metglas sheets, and niobium
foils. Tests carried out in an ADR system in NIST suggest that 
adequate magnetic shielding is fairly straightforward to achieve.
More detailed characterization of the noise needs to be done to 
answer these questions further.

These detectors will first be used in the upgrade of the Robinson gravitational 
wave telescope\cite{yoon06}, and SPIDER, a proposed balloon-borne experiment 
\cite{montroy06}.
Robinson II and SPIDER share common design concepts, including large throughput 
cold refractive telescope that produces extremely low cross-polarization,
instrument polarization, instrument loading, and beam ellipticity. 
The 30 cm compact optics enables detailed pre-flight characterizations,
and provides an angular resolution of $\sim 1^\circ$ suitable for primordial 
B-mode polarization.

Despite these similarities, the two experiments are based on very 
different observing strategies.
SPIDER is targeting very large angular scale CMB polarization, including the 
re-ionization bump at $\ell\sim 8$. It will survey 50\% of the sky.
To facilitate Galactic foreground removal on large angular scales, 
SPIDER payload consists of 6 monochromatic telescopes, covering a wide 
frequency range of 80-275 GHz with 2312 antenna-coupled TES bolometers.
Multiple telescopes enable each to be monochromatic, thereby improving
the performance of AR coatings, enabling a simplified wave plate design,
and providing a large system throughput with minimum aperture size.
On the other hand, Robinson II will be observing from South Pole through the
100 GHz and 150 GHz atmospheric windows. With the long integration time available 
from the ground, it will go extremely deep on 1-2\% of the sky that has minimum 
astronomical foregrounds.  

Both experiments will be an important scientific pathfinder for CMBPol, a 
comprehensive NASA satellite to study the CMB polarization~\cite{weiss05}. 
In addition, many technical aspects discussed in this paper will be 
thoroughly tested. We believe the antenna-coupled TES detector technology is
a strong candidate for CMBPol.
 
\acknowledgments     

The authors acknowledge support from the JPL Research and Technology
Development program for supporting the TES development and SQUID
multiplexer testing, and a 2004 NASA/APRA grant "Antenna-Coupled TES
Bolometer Array for CMB Polarimetry" to J. Bock.
CLK acknowledges the support of a NASA postdoctoral fellowship.


\bibliography{tes}   
\bibliographystyle{spiebib}   

\end{document}